# Finite Difference Based Wave Simulation in Fractured Porous Rocks

Janaki Vamaraju and Mrinal K. Sen

*Department of Geological Sciences*
*The University of Texas at Austin*

## ABSTRACT

Biot's theory provides a framework for computing seismic wavefields in fluid saturated porous media. Here, we implement a velocity stress staggered grid 2D finite difference algorithm to model the wave propagation in poroelastic media. The Biot's equations of motion are formulated using a finite difference algorithm with fourth order accuracy in space and second order accuracy in time. Seismic wave propagation in reservoir rocks is also strongly affected by fractures and faults. We next derive the equivalent media model for fractured porous rocks using the linear slip model and perform numerical simulations in the presence of fractured interfaces. As predicted by Biot's theory a slow compressional wave is observed in the particle velocity snapshots. In the layered model, at the boundary, the slow P-wave converts to a P-wave that travels faster than the slow P-wave. We finally conclude by commenting on the major details of our results.

## INTRODUCTION

Wave propagation in porous media is of interest in various diversified areas of science and engineering. The theory of the phenomenon has been studied extensively in soil mechanics, seismology, acoustics, earthquake engineering, ocean engineering, geophysics, and many other disciplines. Particularly, in reservoir geophysics, ultrasonic and seismic modeling in porous media is used to study the properties of rocks and characterize the seismic response of geological formations. Seismic survey is a powerful tool to find potential locations of new oil and gas reservoirs and to monitor existing reservoirs. Seismic waves contain essential information on subsurface properties. To quantify the dependence between wave propagation characteristics and subsurface properties related to porous medium, various models have been proposed in literature. The commonly used equations for wave propagation in poroelastic solids are Biot's equations (Biot, 1956a, b, 1962). A macroscopic description of fluid-filled porous media was developed by Gassmann (1951) and Biot (1941, 1956). Biot established the fundamental theory for wave propagation in fluid-saturated porous media. It is still well-accepted and forms the basis for wave propagation studies in porous media. The theory successfully predicted the existence of the second compressional wave (the slow compressional wave), which had been observed in the laboratory (Plona 1980; Berryman 1980).

For numerical simulation of wave propagation in general heterogeneous poroelastic media, the finite-difference method is probably the most widely used technique (Zhu and McMechan, 1991; Dai et al., 1995; Jianfeng, 1999; Pride et al., 2004; Masson et al., 2006; Sheen et al., 2006; Masson and Pride, 2007). To introduce the classical PML in this method in the case of elastic media, the wave equation usually is formulated as a first-order system in time based on velocity and stress (Collino and Tsogka, 2001). In this study, we use a velocity-stress finite difference algorithm with PML absorbing boundaries to model the wave propagation in the porous media. This algorithm is fourth order accurate in space and second order accurate in time.

Further, fractured reservoirs have attracted increasing attention in the research of exploration and production geophysics. Further wave studies are needed to reveal the physics of wave propagation in such a fracture-pore system, which is meaningful to the application in exploration activities. Subsurface fractures can vary in size from microcracks to large faults and strongly influence seismic wave propagation, by causing wave scattering and anisotropy (Crampin 1977, 1981). In this report, we adopt an equivalent model (Qizhen et al., 2011) based on the linear slip theory (Nakagawa and Schoenberg, 2007) to describe the fractured porous media at seismic frequency bands and simulate wave propagation in fractured porous media.

In the following sections, we first briefly review Biot's theory of poroelasticity and present the equivalent media model for fractured porous rocks. We then perform numerical simulations both in fractured and non-fractured porous media and show the analysis results. We finally conclude by commenting on the major details of our results.

**THEORY AND METHOD**

**Biot's theory of poroelasticity**

Maurice Biot established the theory of poroelasticity in early fifties. He made the following assumptions to derive the equations of motion in the porous media: (1) the rock frame is assumed to be elastic; (2) the pores are connected so that the fluid could travel through the pore space; (3) the seismic wavelength is much larger than the average pore size; (4) the deformations are so small that the mechanical processes become linear; (5) the medium is statistically isotropic (Zhu and McMechan, 1991). Although Biot extended his theory to more general cases such as anisotropic porous media, in this report we concentrate on the isotropic case. From the stress-strain relation for a porous medium (Biot, 1962), the solid stress and the pore fluid pressure P are given by

$$\tau_{ij} = 2\mu e_{ij} + (\lambda_c e_{kk} + \alpha M \varepsilon_{kk})_{ij} \quad (1)$$
$$P = -\alpha M e_{kk} - M \varepsilon_{kk} \quad (2)$$

where $e_{ij} = \nabla \cdot u = \frac{1}{2}\left(\frac{\partial u_i}{\partial x_j} + \frac{\partial u_j}{\partial x_i}\right)$ is the solid strain, with $u$ being the particle velocity of the solid, and $\varepsilon_{ij} = \nabla \cdot (u - U)$, where $U$ is the particle displacement of the fluid, $\mu$ is the shear modulus and $\lambda_c$ is the Lame parameter of the saturated rock. $\alpha$ is defined by $1 - \frac{K_{Dry}}{K_{Solid}}$ where $K_{Solid}$ and $K_{Dry}$ are the bulk moduli of the solid and the dry rock frame, respectively. $M$ is coupling modulus defined by $\left[\frac{\phi}{K_{Fluid}} + \frac{\alpha-\phi}{K_{Solid}}\right]$ with $K_{Fluid}$ being the bulk modulus of the fluid.

Inserting $w = u - U$, the particle displacement vector of the fluid relative to the solid, $\rho_f$ and $\rho$ are the fluid and saturated rock densities and $b$ is the fluid mobility defined by $\eta/\kappa$, with $\eta$ being the viscosity of the fluid and $\kappa$ being the permeability of the porous rock. $m = T\frac{\rho_f}{\phi}$ is the effective density of the fluid, with T being the tortuosity. Defining the solid and relative particle velocities $V = \frac{\partial u}{\partial t}$ and $W = \frac{\partial w}{\partial t}$ these equations could be written as first order equations of motion for a statistically isotropic porous media saturated with viscous fluid:

$$(m\rho - \rho_f^2)\frac{\partial V_i}{\partial t} = m\frac{\partial \tau_{ij}}{\partial x_j} + \rho_f bW + \rho_f \frac{\partial P}{\partial x_i} \tag{3}$$

$$(m\rho - \rho_f^2)\frac{\partial W_i}{\partial t} = -\rho_f \frac{\partial \tau_{ij}}{\partial x_j} - \rho bW - \rho \frac{\partial P}{\partial x_i} \tag{4}$$

$$\frac{\partial \tau_{ij}}{\partial t} = 2\mu\frac{\partial e_{ij}}{\partial t} + \left(\lambda_c \frac{\partial e_{kk}}{\partial t} + \alpha M \frac{\partial \varepsilon_{kk}}{\partial t}\right)_{ij} \tag{5}$$

$$\frac{\partial P}{\partial t} = -\alpha M \frac{\partial e_{kk}}{\partial t} - M \frac{\partial \varepsilon_{kk}}{\partial t} \tag{6}$$

**Staggered-grid finite difference**

To model the wave propagation in poroelastic media, equations (3) to (6) need to be discretized using the finite difference algorithm. The unknowns are the solid stresses $\tau_{xx}, \tau_{zz}, \tau_{xz}$, the fluid pressure $P$, the solid particle velocities $V_x$ and $V_z$, and the relative fluid particle velocities $W_x$ and $W_x$. For more accuracy, a velocity-stress staggered-grid finite difference scheme was used for numerical modeling. The solid stresses and the fluid pressure are calculated on the regular grid whereas other unknowns are calculated on the staggered grid where the grid points are shifted by half a grid size (Levander, 1988) (Figure 1).

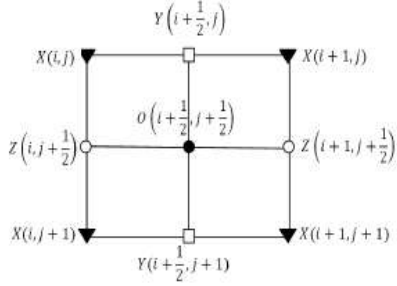

Figure 1:The staggered-grid layout and the locations of stress, velocity and pressure components. Normal stress and pressure represented by X, Horizontal velocities by Y, vertical velocities by Z and shear stress by O.

**Equivalent medium model for fractured porous rocks**

In the Linear slip model (Nakagawa and Schoenberg, 2007), the stress field and the pressure field is assumed to be continuous across the fracture whereas the displacement field is assumed to be discontinuous. The jump in the displacement is linearly related to the traction vector at the fracture as follows:

$$[V_x] = \eta_T \tau_{xz} \tag{7}$$

$$[V_z] = \eta_{Nd}\{\tau_{zz} + \alpha P\} \tag{8}$$

$$[W_z] = -\alpha[V_z] + \eta_M(-P) \tag{9}$$

where $\eta_T$ is the shear compliance, $\eta_{Nd}$ is the dry normal compliance and $\eta_M$ is the fluid/mineral compliance of the fracture. Using this model Q. Du et al., 2011 derived the equivalent medium model for vertical fractures. We shall adopt this model for the simulations in this report.

**RESULTS AND DISCUSSION**

**Homogeneous isotropic model without fractures**

We first simulate the poroelastic wave equations in a homogeneous isotropic medium. A ricker wavelet was used as an explosive source. To inject an explosive, source the wavelet is added to the normal solid stresses and fluid pressure. The size of the model is 2km by 2km. The source is located at the position (x, z) = (1.0,1.0) km with a dominant frequency of 40 Hz. The recorders are located at 600m depth. Grid spacing dx = dz = 2m and time step dt = 0.2ms. The properties of the medium are shown in Table 1. The calculated snapshots for the vertical particle velocities of the solid and relative fluid with respect to solid are show in Figure 2. Two P-waves traveling in the medium can be seen where the slow P-wave travels with a speed close to the wave speed in the fluid. The particle velocity of the slow P-wave in the solid is out of phase with the one in the fluid. In contrast, for the fast P-wave the particle velocities are in phase in the solid and fluid. Figure 3 also shows the fluid pressure snapshots for this model. The slow P-wave has a larger relative amplitude than the fast P-wave which is because the slow P-wave is originated from the fluid movement. Vertical and horizontal velocity shot gathers are also plotted below. As observed from the snapshots the slow P wave can also be seen in the poroelastic shot gather.

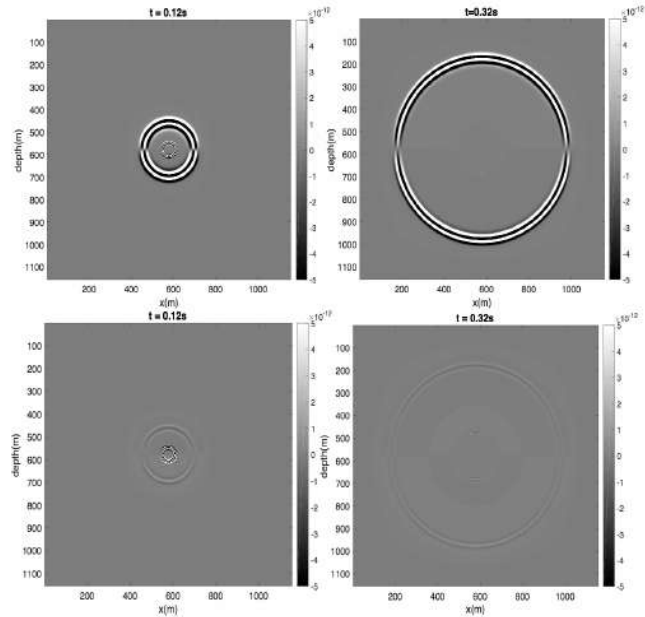

| Property | Units | Value |
|---|---|---|
| Solid density $\rho_s$ | kg/m³ | 2588 |
| Fluid density $\rho_f$ | kg/m³ | 952.4 |
| Matrix tortuosity a | | 2.49 |
| Porosity $\phi$ | | 0.25 |
| Bulk density $\rho = \varphi\rho_f + (1-\varphi)\rho_s$ | kg/m³ | 2179.1 |
| Apparent density $\rho_w = a\rho_f/\varphi$ | kg/m³ | 9486 |
| $\alpha$ | | 0.89 |
| M | Pa | $7.71 \times 10^9$ |
| Damping viscous term K | Ns/m⁴ | $3.38 \times 10^5$ |
| Fast pressure-wave velocity in the solid $V_{Pf}$ | m/s | 2817.33 |
| Slow pressure-wave velocity in the solid $V_{Ps}$ | m/s | 740 |
| Shear-wave velocity in the solid $V_S$ | m/s | 1587.4 |
| Shear modulus $\mu$ | m/s | $5.25 \times 10^9$ |
| Lamé coefficient in solid matrix $\lambda_s$ | Pa | $6.2 \times 10^8$ |
| Lamé coefficient in saturated medium $\lambda = \lambda_s + M\alpha^2$ | Pa | $6.7271 \times 10^9$ |

Table 1: Properties of the homogeneous medium

Figure 2: Vertical velocity snapshots of the solid (top) and relative fluid with respect to the solid (bottom) at t=012s, 0.32s

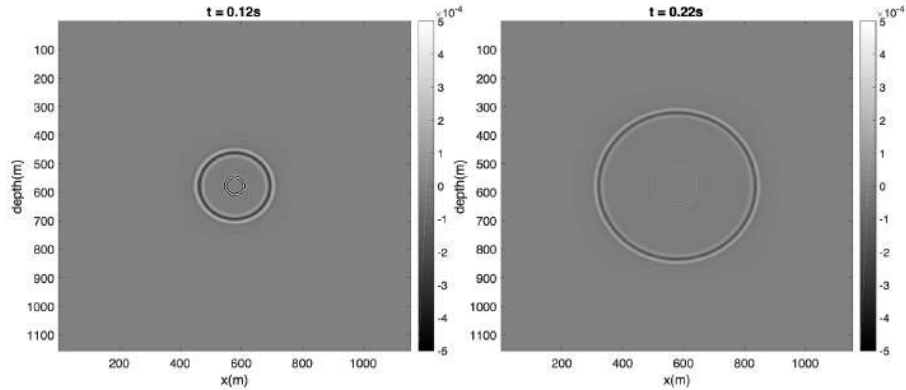

Figure 3: Pressure field snapshots at t=0.12s and 0.22s

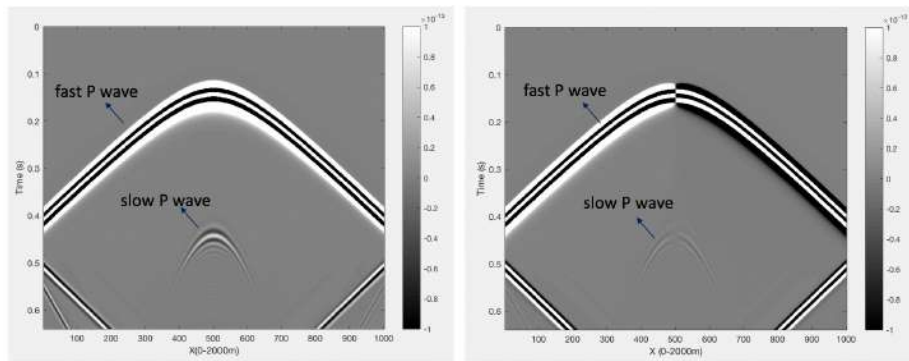

Figure 4: Vertical and horizontal velocity traces for the homogeneous medium.

**Homogeneous isotropic model with a vertical fault**

We repeat the previous experiment with a vertical fault line at x = 1km. The source is located at the position (x, z) = (0.5, 0.5) km with a dominant frequency of 40 Hz. The recorders are located at 100m depth. The fault has the properties shown below. Snapshots and shot gathers similar to the previous example are shown in figures 5-6. The results show that the linear-slip discontinuity generates reflected and transmitted fast P- and S-waves with their corresponding head waves. The reflected slow wave dissipates very quickly and hence cannot be seen.

Material properties (Berea sandstone)

$K_s = 39$ GPa, $E = 14.4$ GPa, $\nu = 0.2$, $\rho_s = 2500$ kg/m$^3$, $k_m = 1 \times 10^{-12}$ m$^2$, $\phi_m = 0.178$, $\phi_f = 0.0178$

Fluid properties (water)

$K_f = 2.3$ GPa, $\mu^f = 0$ GPa, $\eta^f = 0.001$ Pa S, $\eta = 0.001$ Pa S, $\rho_f = 1000$ kg/m$^3$, $k_f = 1 \times 10^{-6}$ m$^2$

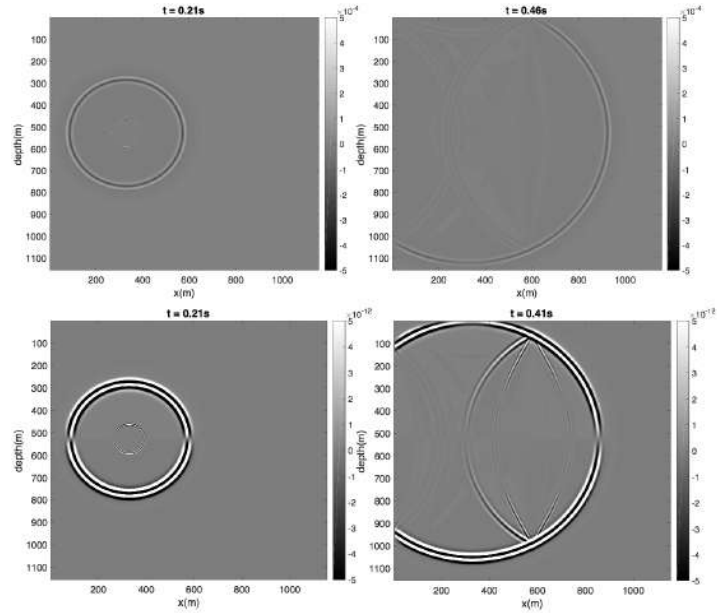

Figure 5: Pressure field and vertical solid velocity snapshots at t=0.21 and t=0.46s

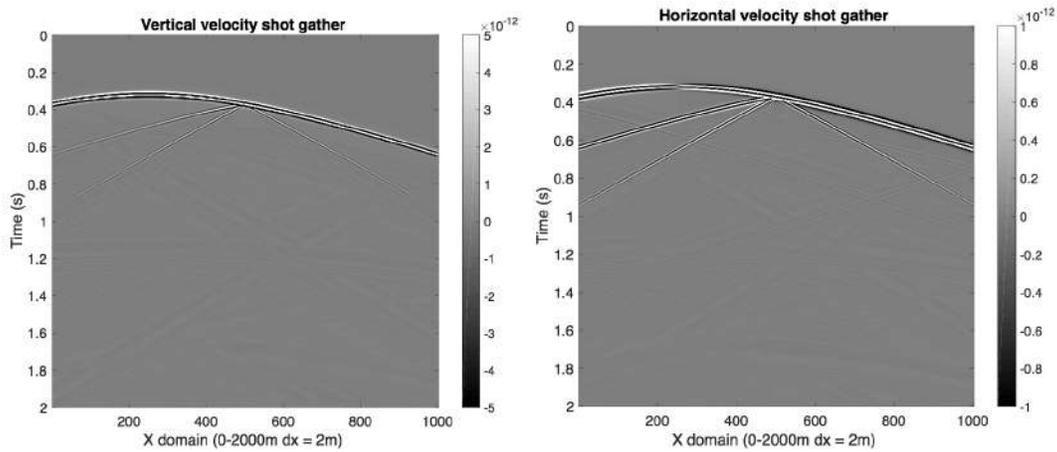

Figure 6: Horizontal and vertical velocity traces for homogeneous medium with vertical fault line

**Heterogeneous 2-layer model**

The next example we use to study is a 2-layer heterogeneous model. Properties of the first layer are given in Table 2. For the second layer, we use the same properties as in the first example for a homogeneous medium. Both the layers have 1km depth. The source is located at the position (x, z) = (1.0, 0.75) km with a dominant frequency of 40 Hz. The recorders are located at 500m depth. Particle velocity snapshot is shown in Figure 7. The shot gather is also plotted in Figure 8.

We see several wave modes including the converted waves. $P_f$ and $P_S$ are the fast and the slow P-waves, respectively. $P_{ff}$ is the reflected fast P-wave, $P_{SS}$ is the reflected slow P-wave, and $P_{fS}$ is the reflected S-wave that is converted from a fast P wave. A reflected fast P-wave which is converted from the slow P-wave is also evident in this figure. This wave which is marked as $P_{Sf}$, travels faster than the reflected slow P-wave.

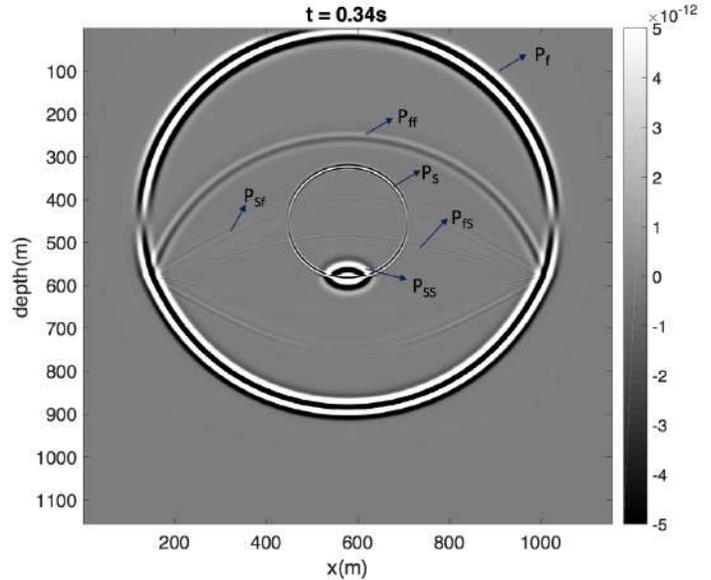

| | | |
|---|---|---|
| Solid density | kg/m³ | 2400 |
| Fluid density | kg/m³ | 1000 |
| Porosity | | 0.2 |
| Bulk density | kg/m³ | 2120 |
| Apparent density | kg/m³ | 15000 |
| $\alpha$ | | 0.2 |
| M | Pa | 11.25e9 |
| Damping viscous term | Ns/m⁴ | 0 |
| Shear modulus | Pa | 5.4e9 |
| Tortuosity | | 3 |
| Lame coefficient in saturated medium | Pa | 7.15e9 |

Table 2: Properties of the upper layer in the 2-layer model

Figure 7: Vertical solid velocity snapshot at t=0.34s

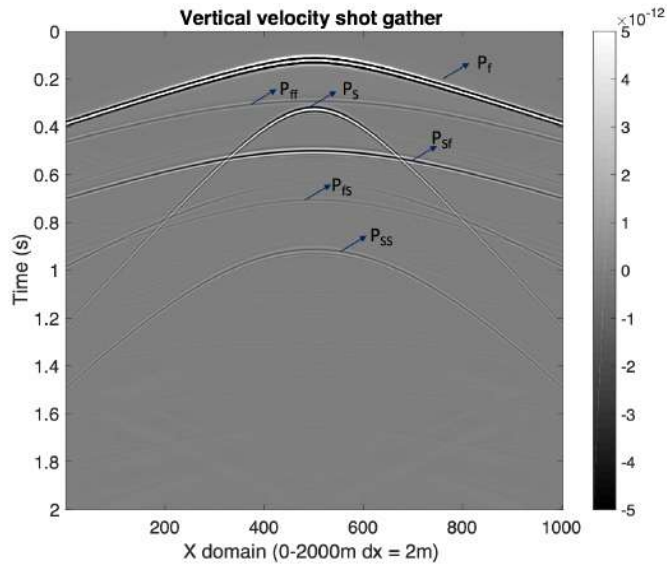

Figure 8: Vertical velocity shot gather for the 2-layer model.

## CONCLUSIONS

Biot's poroelastic wave equations were modeling using a velocity stress staggered grid finite difference algorithm. We performed simulations both in fractured and non-fractured media using the linear slip theory in the former case. As predicted by Biot's theory, two compressional waves appeared in the snapshots of the particle velocities along with the converted waves due to the fault line. A 2-layered model was also simulated to see the wave behavior at a boundary. The vertical velocity snapshots showed different wave modes generated at the interface including a fast P-wave which is converted from the slow P-wave. This poroelastic FD algorithm has the potential to be used for full waveform inversion to obtain porous media properties that are ignored in elastic wave equation algorithms.